\documentclass[twocolumn]{IEEEtran}
\usepackage{mathptmx}  
\usepackage{amsmath}
\usepackage{amsthm}
\usepackage{amsfonts}
\usepackage{amssymb}
\usepackage{graphicx}
\usepackage{url}
\usepackage{color}    
\usepackage{cite}
\usepackage[ruled]{algorithm2e}
\usepackage[bookmarks=false]{hyperref}
\usepackage{tabu}
\usepackage{lmodern}
\usepackage[noend]{algpseudocode}
\usepackage{indentfirst}
\usepackage{xspace}
\usepackage{booktabs}
\usepackage{wasysym}
\usepackage{textcomp}
\usepackage{paralist}
\usepackage{multirow}
\usepackage{array}
\newcommand{\subparagraph}{} 
\usepackage[compact]{titlesec}
\usepackage{threeparttable}
\usepackage{adjustbox}
\usepackage[font={footnotesize}]{caption}

\newcommand{\myparagraph}[1]{\noindent\textbf{#1.}\xspace}

\setdefaultleftmargin{1em}{1em}{}{}{}{}

\title{Permissionless Blockchains and Secure Logging}
\vspace{-0.2cm}
\author{Chunpeng Ge$^{1}$, Siwei Sun$^{2,*}$, and Pawel Szalachowski$^{1}$ \\
	\vspace{-0.1cm}
    \small{$^1$\textit{ST Electronics - SUTD Cyber Security Laboratory, Singapore University of Technology and Design,} Singapore\\\vspace{-0.2cm}
    	$^2$\textit{Data Assurance and Communication Security Research Center,
        CAS,} China}
\vspace{-0.7cm}
    \thanks{This research was supported by ST Electronics and National Research Foundation (NRF), Prime Minister's Office Singapore, under Corporate Laboratory @ University Scheme (Programme Title: STEE
Infosec - SUTD Corporate Laboratory).

$^*$This work was done while the author was at SUTD.}
}

\begin{document}
\maketitle

\begin{abstract}
The blockchain technology enables mutually untrusting participants to reach consensus
on the state of a distributed and decentralized ledger (called a blockchain)
in a permissionless setting.  The consensus protocol of the blockchain
imposes a unified view of the system state over the global network, and once
a block is stable in the blockchain, its data is visible to all users
and cannot be retrospectively modified or removed.  Due to these properties,
the blockchain technology is regarded as a general consensus
infrastructure and based on which a variety of systems have been built.

This article presents a study and survey of permissionless blockchain systems in
the context of secure logging.  We postulate the most essential properties
required by a secure logging system and by considering a wide range of
applications, we give insights into how the blockchain technology matches
these requirements.  Based on the survey, we motivate related research
perspectives and challenges for blockchain-based secure logging systems, and
we highlight potential solutions to some specific problems.

\end{abstract}

\section{Introduction and Background}
\label{sec:intro}
\label{sec:background}
\myparagraph{Secure Logging}
Logging is indispensable for many secure IT systems.  While there is no
unanimous agreement on the definition of a secure logging system, it can be
regarded as a database system which securely keeps track of records for
security-critical data.  A secure logging scheme has a wide
range of applications.  It can be a stand-alone system or an integral part of a
larger system.

An \textit{event logging system} is one traditional
application of secure logging.  It records system events of forensic value in a
protected database.
Such logging systems are security-critical regular
targets of sophisticated attackers trying to eliminate their footprints.
Therefore, it is important to prevent unauthorized modifications and deletions
of the log entries.
\textit{Timestamping service} is an infrastructure used to prove the existence
of certain digital data prior to a specific point in time.  It is important to
guarantee the accuracy and validity of the timing information of data and
events, since its defect may have significant security and financial
implications.  Therefore, the accuracy and immutability of the timestamps are
essential.

The security of many systems today is
bootstrapped from securely obtaining
some specific authoritative information.
For example, a PKI is meaningless if the users relying on
it cannot obtain the correct certificates
in the first place~\cite{arthur2011rogue,roberts2011phony}.
Another example are trusted directory servers~\cite{tor2004}, which when
attacked can compromise properties of a relying infrastructure~\cite{attackTor}.
To address these issues, \textit{transparency
logs}~\cite{laurie2013certificate,dowling2016secure,chase2016transparency,TP_log}
have been proposed,
which are services securely maintaining a list or dictionary of objects.
To prevent malicious entries from being inserted into the log without
being noticed, the dictionary should be append-only.
Moreover, participants should have a singleton view of the dictionary, i.e.,
the log should not be able to equivocate -- this usually requires
a gossip protocol to be deployed~\cite{chuat2015efficient,nordberg2015gossiping}.

\myparagraph{Desired Properties}
From the example applications presented,
we can extract a list of desired properties:

	\noindent{\it Availability:}
	The logger can log artifacts
	without significant delays. For clients relying on the
	log server, all logged artifacts and events can be accessed.
	{\it Authenticity:}
	It should be verifiable who has created or submitted the logged artifacts.
	{\it Immutability:}
	Once an artifact has been logged, it cannot be altered or removed without being noticed.
	{\it Non-equivocation:}
	All system participants should have a unified view of the logs.
	The log server cannot present different views of the log for different users.
	{\it Freshness:}
	Some applications may demand the freshness property,
	which allows to order the logged artifacts (weak freshness) or even
	to determine the exact time of them up to certain precision (strong freshness).
	




\myparagraph{Blockchain and Secure Logging}
Blockchain technologies, like Bitcoin~\cite{nakamoto2008bitcoin} and
Ethereum~\cite{wood2014ethereum}, are successful beyond all expectations. This
success is mainly driven by their properties: \textit{consensus}: all parties
can (eventually) agree on the current state of the system,
\textit{transparency}: all transactions (of all participants) are visible to
anyone, \textit{irreversibility}: blockchains have the append-only property
which implies that whenever a transaction is appended to a blockchain it cannot
be retrospectively modified or removed, \textit{decentralization and openness}:
everyone can participate in the system, and no centralized entity authorizes
participants or their transactions, \textit{availability}: the infrastructure is
robust as it can tolerate a large fraction of faulty participants.
Due to these properties, the blockchain technology enables novel applications
like cryptocurrencies and smart contracts. Even now, shortly after their
advents, these systems are successful and as a consequence of this success
developers and researchers try to reuse blockchain infrastructures to build new
or enhance existing systems.

A secure logging service based on decentralized blockchain technology could have
great potential and could be deployed by multiple existing applications and used
for empowering novel ones.
In fact, there are proposals that try to use blockchain as a logging-related
service. For instance, blockchain-based timestamping~\cite{ots}, trusted record-keeping
service~\cite{gao2017decentralized}, decentralized audit systems~\cite{liwill},
document signing infrastructures~\cite{jamthagen2016blockchain}, timestamped
commitments~\cite{Clark12commitcoin}, or secure off-line payment
systems~\cite{dmitrienko2017secure}. 
Another line of research in this area is to design transparency schemes based on
blockchain technologies, such as key transparency~\cite{bonneau2016ethiks},
certificate transparency~\cite{matsumoto2016ikp,szalachowski2018blockchain},
binary transparency~\cite{contour}, or log transparency~\cite{tomescu2017catena}.
Other related work includes providing legacy content (e.g., web content) to
smart contracts~\cite{zhangCCS16,ritzdorf2018tls,juan_PDFS}.

However, there are many challenges  associated with
designing and deploying such systems.
In this work, we study these systems, their logging-relevant properties, show
their limitations, and research opportunities.

\section{Selected Blockchain Platforms}
\label{sec:protocols}
\begin{table*}[t!]
	\renewcommand\arraystretch{1.1}
	\centering
	
	\caption{Logging-Related Features of Selected Platforms.}
    \vspace{-0.1cm}
	\begin{tabular*}{0.9\textwidth}{@{\extracolsep{\fill}} lccccccc}
		
		&
		tx arrival&
		public-key&
		publicly&
		data&
		timestamp&
		data size&
		data\\
		\textit{Platforms} &
		time&
		identities&
		accessible&
		structure&
		range&
		per tx&
		recording\\
		\toprule
		
		Bitcoin   &10 min             &yes              &yes       &chain     &2 h                &220 Bytes   &\texttt{OP\_RETURN}\\
		Ethereum  &15 sec          &yes              &yes      &chain     &15 s                  & 780 KBytes         &smart contract\\
		IOTA      & net. latency         &yes              &yes      &DAG       &$\perp$            &1.27 KBytes  &message\\
		\bottomrule
	\end{tabular*}
	\begin{tablenotes}
		\footnotesize
        \centering
		\item[*] Note that, in IOTA, there is no validity check of the
            timestamp; thus it can be arbitrary and we use $\perp$ to represent it.
		
	\end{tablenotes}
	
	\label{Table 1:}
    \vspace{-0.4cm}
\end{table*}
\myparagraph{Bitcoin} Bitcoin~\cite{nakamoto2008bitcoin} is the first and largest
cryptocurrency and due to its open, distributed, decentralized nature, and use
of public-key cryptography, it offers
(transaction) authenticity and a certain degree of availability.
The Bitcoin network maintains a distributed and replicated ledger (i.e.,
blockchain) --
an append-only linked list of blocks (containing transactions).
Since the
system is permissionless, any participant can vote her own view of
the current state by trying to append new blocks to the blockchain.
To combat Sybil attacks and reach an agreement on the system state
across the network, Bitcoin employs the Nakamoto consensus where
a solution of a computational puzzle,
serving as a Proof-of-Work (PoW) must be presented to append a new block
to the blockchain. An incentive structure is embedded in the protocol to
encourage participants constantly competing to put their own blocks
onto the blockchain.

The global ledger is append-only,
and once a block is stable in the blockchain, its data cannot be retrospectively
modified or removed without significant computational resources.
Moreover, the whole network has a unified view of the blockchain.
These properties lead to a natural way to build a secure logging system
providing the non-equivocation property, where we can record the log statements on
the blockchain. This can be done by sending special transactions
in the Bitcoin network. For example, the \texttt{OP\_RETURN} code allows 
adding 220 Bytes of arbitrary data to a transaction output.

Since every block of the Bitcoin blockchain has a timestamp, when recording data on the
blockchain, it may be tempted to use the same timestamp for the data.
In practice, timestamps can differ radically from
the actual time, and they are susceptible to
manipulation~\cite{gervais2015tampering,timeJacking,szalachowski2018short}.
Hence, the accurate time cannot be determined and extra caution must be taken when
using the Bitcoin timestamps.
Bitcoin introduces the \textit{unspent transaction output} (UTXO) model where
new transactions can spend only UTXOs (i.e., actual coins) included in existing
transactions.
Bitcoin introduces light \textit{SPV clients} which can interact with the
blockchain without possessing and validating all blocks (they store and validate only short
block headers).

\myparagraph{Ethereum} Ethereum~\cite{buterin2017ethereum} is
a decentralized and open replicated state machine whose state is maintained as a
PoW blockchain.
Ethereum keeps track of a general-purpose state which
can be represented as a global dictionary comprised by key-value pairs.
The state transition of Ethereum is processed by the so-called
{\it Ethereum virtual machine}
executing code (called {\it smart contract})
written in a Turing complete language.
Ethereum introduces a native cryptocurrency called
{\it ether} and the notion of {\it gas}.
Ether is not only an integral
part of the underlying PoW based blockchain, but also intended as
a utility currency to purchase the gas that will be consumed
when using the system resources.
This provides economic incentives and security to the system.

Since Ethereum uses a similar consensus mechanism as Bitcoin,
any secure logging systems implemented based on Bitcoin can also be
realized over Ethereum, and they can achieve similar properties
with respect to availability, authenticity, immutability and non-equivocation.
Moreover, with Ethereum
one can implement smart contracts with almost arbitrary logic.
Thus, compared to Bitcoin, Ethereum is a more suitable choice
if a logging service requires actions or computations
to be executed automatically according to the current state and user inputs.
Finally, Ethereum provides a better freshness property than Bitcoin, however,
nodes of the Ethereum network
rely on the NTP~\cite{mills1985network} servers, and
therefore their timestamps are generated in
a centralized way to some extent.

\myparagraph{IOTA}
There are multiple proposals aiming to improve the efficiency of blockchain-based
systems by deploying  directed acyclic graph (DAG)
instead~\cite{popov2016tangle,sompolinsky2016spectre,sompolinsky2018phantom,sompolinsky2015secure,li2018scaling}.
IOTA~\cite{popov2016tangle} is a permissionless distributed ledger where
transactions are stored in
a data structure whose logical topology forms a
DAG.
This design aspires to resolve some inherent
scalability issues of chain style
blockchain and
positions itself as suitable for IoT applications.

In the IOTA terminology, the {\it tangle} is the data structure storing the
distributed ledger, whose vertices are called {\it sites}.
Each {\it site} contains one transaction issued by the
IOTA user network. To be permanently attached to the
{\it tangle} and become one {\it site}, a transaction must directly
approve two existing transactions ({\it sites}) in the tangle.
%
If there is a path from {\it site} $B$ to {\it site} $A$,
we say that {\it site} $A$ is indirectly approved by
{\it site} $B$. The {\it genesis site} is directly or
indirectly approved by all {\it sites} (excluding itself)
in the {\it tangle}. The {\it tips} are those {\it sites}
that have not been approved by any {\it site}.
Consequently, the chronological order of two {\it sites}
cannot be determined unless there is a path
connecting them. Thus even the weak freshness on different paths cannot be determined.
In IOTA, anyone can issue a data transaction with arbitrary content
of about 1.27~KBytes.
Though each transaction has a timestamp field, it is not verified when the transaction is
added to the IOTA network which means this timestamp can be any time with the correct format.
Therefore, it is challenging to build time-sensitive logging systems relying
only upon IOTA.
Currently, the security of the IOTA network
is ensured by an entity called {\it coordinator} who
verifies all transactions,
that is, a transaction cannot be a
part of the {\it tangle} without the coordinator's approval.
Consequently, the community calls into question the (de)centralization nature of
IOTA, and we do not find any convincing response from the IOTA Foundation.

\section{Selected Blockchain-based Logging Systems}
\label{sec:systems}


\begin{figure*}[ht!]
	\centering
	\includegraphics[width=0.75\linewidth]{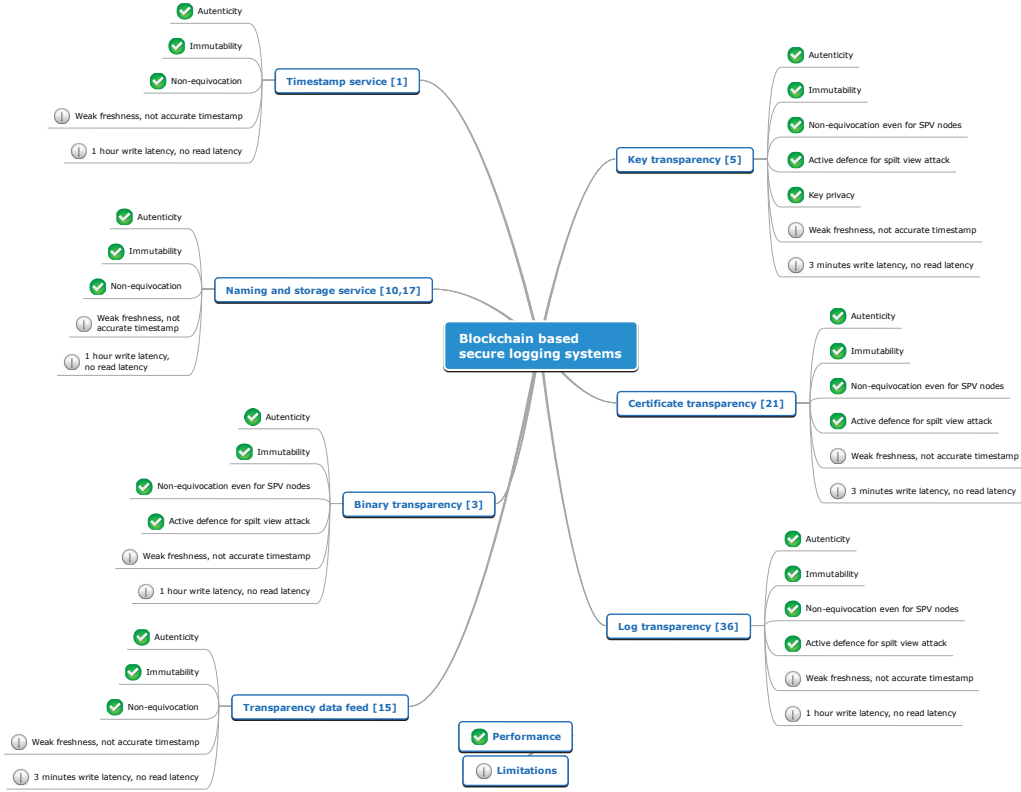}
	\caption{Categorization of blockchain-based secure logging systems.}
	\label{fig:Blockchain based secure logging systems}
\end{figure*}

\myparagraph{Namecoin}
Namecoin is a decentralized key-value pairing
log system based on a Bitcoin
hard fork~\cite{namecoin} preserving its main properties. Namecoin achieves human-readability, strong ownership and
decentralization for a naming log system while no previous
systems can provide both these three properties.
In Namecoin, a user registers a key-value record
on the blockchain by issuing a special transaction containing the record.
Once this transaction is included in
the blockchain,
the record creation operation is done. This record and owner address
will be seen by every node in the blockchain network.
For updating the record, the owner issues a transaction
containing the updated information.
The initial motivation for Namecoin was to create an
alternative to DNS.
The latency of creating and updating records is capped by the Bitcoin's consensus protocol,
and its average time is 60 minutes.
The authentication property is achieved by a pseud-anonymous address as its identity.
For freshness, Namecoin can prove the order of the
name-value records. However, the exact time of a record cannot be guaranteed.

\myparagraph{Commitcoin}
Commitcoin~\cite{Clark12commitcoin} is a timestamped commitment scheme
based on Bitcoin. When the commitment is opened, anyone can be convinced that the commitment
was made before a certain time.
Assume that Alice is a Bitcoin user with a key pair $(sk, pk)$ who wants to
make a commitment of message $m$. Alice first computes the commitment
$c$ of the message $m$ with random number $r$, and then derives a new key pair $(sk', pk')$ with the private
key $sk' = c$. Then Alice signs a Bitcoin transaction $\tau_1$
which sends 2 bitcoins from $pk$ to $pk'$ with secret key $sk$ and randomness
$\rho$, producing signature $\sigma_1$.
Alice signs another transaction $\tau_2$ which send $1$ bitcoin from $pk'$ to $pk$ with
secret key $sk'$ and
randomness $\rho'$, producing signature $\sigma_2$.
The signed transactions are broadcast to the Bitcoin network to be included
in the public blockchain, which proves that Alice knows the corresponding private
keys of $pk$ and $pk'$. Alice can make the commitment publicly available by
signing a transaction $\tau_3$ which returns the remaining
1 bitcoin from $pk'$ back to $pk$ with secret key $sk'$ and
previously used randomness $\rho'$ and broadcasting the resulting signature $\sigma_3$ to the Bitcoin network.
Note that this operation effectively leaks $sk' = c$ to the public since the
same key and randomness are used to generate the signatures $\sigma_2$ and $\sigma_3$~\cite{Clark12commitcoin}.
Finally, Alice can open the commitment by announcing $(m, r)$, and the timestamp of the
block containing $\tau_1$ indicates a rough time at which the commitment was created.
The accuracy of commitment timestamps depends on Bitcoin timestamps.

\myparagraph{Catena}
Catena~\cite{tomescu2017catena} is an efficient non-equivocation scheme built on top of Bitcoin.
A Catena log is bootstrapped by issuing an initial transaction
to the Bitcoin blockchain called the {\it genesis transaction}.
To issue the first statement in the log associated with a genesis transaction,
Catena commits the statement $s_1$ via an \texttt{OP\_RETURN} transaction whose input
is the UTXO of the genesis block. Similarly, any subsequent log statement $s_{i+1}$ is
embedded in an \texttt{OP\_RETURN} transaction that spends the UTXO of $s_i$,
creating a chain of transactions with log statements rooted at the genesis transaction.
The statements are verified against the genesis block.
The resistance  against equivocation is as strong as that of
Bitcoin, since inconsistent statement chains
imply a double spending at some point of the chain.
Catena is an example of an  application
inheriting the security of the underlying blockchain.


\myparagraph{Contour}
Contour~\cite{contour} presents a proactive mechanism for binary transparency.
Contour is built on top of the Bitcoin blockchain. Whenever the authority
wants to issue a package, it incorporates the hash value of each binary as a
leaf of a Merkle tree with root $h_{b}$. Once the Merkle tree reaches a threshold size,
the authority issues a blockchain transaction $tx$ in which $h_{b}$ is embedded
as one of the output by using \texttt{OP\_RETURN}. Like in Catena~\cite{tomescu2017catena},
every such transaction $tx$ must spend a previous transaction output that is
spent by the authority.
When a client requests a software updating, accompanying with the requested binary,
two inclusion proofs which assert the binary has been added in the log and is thus accessible to
the monitor are sent to the auditor. The proofs convince the auditor that a) the relevant binary
is included in the Merkle tree represented by $h_{b}$ and b)
the transaction $tx$ is included in the block.
The authority cannot mutate nor equivocate a published binary
as long as the Bitcoin platform is secure.

\myparagraph{Data Feed for Smart Contracts}
Data feeds for smart contract make off-chain data available for
on-chain smart-contract-based applications.
Town Crier \cite{zhangCCS16} relies on a trusted execution environment (TEE) to
implement a service which contacts a content provider, verifies and parses its
data, and provides it to a smart contract on demand. It does not involve the
content provider in the protocol, however, it requires trust in the TEE platform
used.
TLS-N~\cite{ritzdorf2018tls} provides a transport-layer approach, where content
providers can provide non-repudiation for their application-layer data (e.g.,
HTTP). It is a more general solution, however, it requires low-level protocol
changes and content providers must deploy the protocol.
PDFS~\cite{juan_PDFS} is an application-layer solution giving content providers
smart contracts used to verify the authenticity of their published content.  In
PDFS, off-chain data is obtained from a content provider's website and its
identity is authenticated by a TLS certificate.  The scheme provides a payment
framework, non-equivocation, and censorship evidence for content providers but
it requires them to deploy (only
application-level changes are required).



\section{Research Perspectives and Challenges}
\label{sec:perspectives}
\myparagraph{Reliable Timestamps}
Bitcoin timestamps may be inaccurate. Thus, it is a valuable research topic
to investigate how to enhance the Bitcoin protocol with
existing trusted timestamping services, which can provide evidence that a
block is created within a sharper time interval.
One possible solution is that we can combine the timestamp protocol~\cite{rfc3161} with the blockchain platforms as previously
presented~\cite{szalachowski2018short}.  The main idea is that one
can issue transactions with timestamp authority's timestamped and
signed messages containing references to known blocks of the blockchain. Then
the time interval in which a given block between two blocks containing
timestamped messages can be derived according to the order of the blocks. That
is, we insert anchor points with more accurate timing information into the
blockchain.
A similar idea can be applied to DAG-based systems like IOTA. One can insert anchor
points with reliable timing information and pointers to existing
sites. However, this approach requires not only anchor points but
also weak freshness, which is not provided by IOTA.
Consequently, to what extent we can improve the freshness property
of IOTA is probabilistic in nature which deserves further investigation.

\myparagraph{Cryptographic Data Structures} Currently, most blockchain
technologies such as Bitcoin and Ethereum
attain their security properties in a decentralized way at the cost
of highly redundant and replicated data and computation.
However, storing all logged data on-chain may be impractical, expensive, or
undesired (for privacy issues), and
this issue calls for efficient cryptographic data structures
securely binding on-chain and off-chain data that ideally fulfill the
following properties:
    a) the data structure can produce a
	``digest'' with a fairly small size from the ever-increasing log entries.
    b) From the cryptographic data structure, the log server can efficiently generate
	compact proofs with rich semantics (e.g., append-only proof, (non)membership of
	objects).
    c) The proofs can be verified by clients efficiently.
    d) The blockchain transaction model implies that any data
    on-chain is publicly accessible. Therefore, it is desirable if the
    cryptographic data structure facilitates the implementation of privacy and
    access control policies in the system.


%
%
%
%
%
%
%
%

\section{Conclusions}
\label{sec:conclusions}
We conduct a study and survey of
secure logging systems based on blockchain technologies.
The essential
properties for secure logging systems are identified and
by concrete examples, we show how the blockchain technology
is leveraged to fulfill these requirements.
We also identify several deficiencies
of current systems, and make an initial attempt to solve them.
We signal further research that is needed to better understand and resolve
these deficiencies.


\bibliographystyle{IEEEtranS}
\bibliography{ref}
\end{document}